# Attosecond shaping of high-current pulsed electron beams in a home-built 37-keV beamline

Yuichi Tachibana[1*†], Marie Ouillé[1‡], Takuya Hosobata[1], Takashi Isoshima[1], Yoshiyuki Takizawa[1], Yutaka Yamagata[1] and Yuya Morimoto[1,2*]

[1]*RIKEN Center for Advanced Photonics (RAP), RIKEN, 2-1 Hirosawa, Wako, Saitama 351-0198, Japan*

[2]*RIKEN Pioneering Research Institute (PRI), RIKEN, 2-1 Hirosawa, Wako, Saitama 351-0198, Japan*

*yuichi.tachibana@rs.tus.ac.jp; yuya.morimoto@riken.jp

[†]Current address: Department of Physics, Tokyo University of Science, 1-3 Kagurazaka, Shinjuku, Tokyo 162-8601, Japan

[‡]Current address: Université Paris-Saclay, Institut d’Optique Graduate School, CNRS, Laboratoire Charles Fabry, 91127 Palaiseau, France

**Ultrashort electron pulses with a high average current provide a powerful means of enhancing time-resolved imaging and photon generation. In this study, we report the attosecond shaping of sub-relativistic electron beams using membranes in a newly developed apparatus that delivers a relatively high current (>2 electrons per pulse on a sample) with negligible space-charge effects. Optimizing the membrane arrangement minimizes the spread of electron–light delays to within a femtosecond over a wide range of incident angles. This enables the recording of attosecond streaking spectrograms, where net acceleration and deceleration, as well as monochromatization and energy broadening, are clearly observed. Through comparison with models, we estimate the durations of the bunched electrons to be 1.3 fs (FWHM) and 0.5 fs (RMS). Furthermore, we demonstrate the attosecond modulation of pulsed beams with a large energy spread originating from space charge effects. A modulation amplitude of 2 eV is shown to be sufficient to shape a beam with an initial spread of 15 eV (FWHM). These results represent a significant step toward the generation of an attosecond pulse containing one or more electrons.**

## Introduction

The modulation of electron beams with light [1–14] allows for temporal shaping at frequencies exceeding those of previously available technologies using microwaves by several orders of magnitude, leading to the generation of electron bunches as short as attoseconds [15–22]. Combined with electron microscopy, the optical electron-beam shaping has enabled the observation of optical fields at their natural temporal and spatial scales [23,24] with enhanced image contrast [25]. The ability to coherently modulate the wavefunction of high-energy beams has also paved the way for quantum photon emission [26–32], miniaturized electron accelerators [33,34], diffractive imaging of charges in motion [35–39], and coherent control of electron-matter interactions [40–42].

Attosecond electron pulses are frequently produced in electron microscopes [18,19, 21,23] in which a substantial fraction of the beam is truncated to achieve high spatial resolution. In contrast, maintaining a high flux is crucial for detecting subtle signal changes in certain applications such as electron diffraction from a homogeneous sample and photon generation from a macroscopic target where high spatial resolution is not the primary concern. However, the attosecond shaping of high-flux electron beams requires a homogenous, phase-coherent electron-light interaction over the extended beam diameter, focusing angle, or both. Additionally, the energy spread associated with the space charge effect can also limit the available current for optical shaping.

In this study, we report the development of an electron beamline with a maximal energy of 37-keV and the production of electron pulses with sub-optical-cycle durations (0.5 fs in RMS). The beamline provides a flux of up to ~3 electrons/pulse at the sample location without significant space charge effects (<4 eV FWHM). With a home-built electron-energy spectrometer, we perform a characterization of the optically modulated electron beams with attosecond precision using membranes properly aligned to minimize the electron-light temporal mismatch caused by the beam focusing angle. We also report on the sub-cycle signature appearing in a beam with a significant space charge effect (15 eV in FWHM) that is bunched with a modulation amplitude of merely 2 eV.

## Experiment

The schematic of the experiment is shown in Fig. 1a. The output of the femtosecond laser (1033 nm, 190 fs, 200 kHz) is split and used for both the electron pulse generation and the attosecond shaping. Electron pulses are produced through two-photon emission from a tungsten <310> needle tip (Denka) whose apex was rounded to a few micrometers by discharges. The pulses are accelerated to an energy of 37 keV (velocity $v_e = 0.36c$) by a static field and guided by magnetic deflectors (not shown) and magnetic lenses to the chamber where the beam is modulated by the laser light. The beam is not clipped by any pinholes or apertures after exiting the gun, allowing a high current delivered to the modulation point. The electron flux is monitored by a Faraday cup. For the energy analysis of the beam, we developed a magnetic-prism with a 90-degree deflection angle at a radius of 100 mm. The energy-dispersed electrons are projected onto a microchannel detector using an octupole lens. A movable slit of with dimensions of 0.125 mm×1 mm (~9% transmission) is installed before the modulation point to limit the focusing angle in one dimension ($x$) and improve the energy resolution. We achieve a resolution of 3.2 eV (FWHM) for the 37-keV beam and at a 2-s exposure. The resolution at a 5-keV energy is 1.2 eV (FWHM), which sets the upper limit of the beam's bandwidth in the absence of space charge effects. The duration of the electron pulse is 290 fs (FWHM), as characterized by energy streaking with the 190-fs pulse. Figure 1e shows the relationship between the energy bandwidth and the number of electrons at the sample position without the slit. No significant broadening is observed up to ~3 electrons/pulse. Unless otherwise specified, the experiment is performed with a current of ~2 electrons/pulse (without the slit).

The laser pulses used for the attosecond shaping are stretched to a duration of ~1.8 ps (FWHM) for the temporally homogeneous interaction. Two stretched pulses with a controlled delay time $\Delta t$ are used for the modulation and characterization, respectively. The electron and laser beams intersect at right angles. Two 20-nm-thick silicon-nitride foils (Norcada) facilitate electron-photon coupling [2,17,43,44] without the need of tight beam focusing due to its uniform thickness over the large area. At the first foil (foil-1), the laser electric field induces a temporally periodic energy modulation. After free space propagation over a 2.6-mm distance between the foils ($L$ in Fig. 1b), the temporal density of the beam is modulated, and a train of short pulses is formed. There are 84 micro-pulses separated by one optical cycle (3.44 fs) in the 290-fs envelope. The temporal structure is analyzed through the energy modulation induced by the second foil (foil-2). The electron beam is spatially focused on foil-2 with a spot size ($1/e^2$ full width) of 12 μm ($y$) and 16 μm ($x$). The corresponding focusing angles are $|\alpha_{xz}| = 0.3$ mrad and $|\alpha_{yz}| = 2.3$ mrad. The transmittance of each foil is approximately 40%, resulting in a combined transmittance of 17%. To achieve the spatially homogeneous electron-light coupling and to avoid the Gouy phase shift, we set the laser focal points behind the foils to obtain large spots (140 μm ($x$) and 300 μm ($y$), $1/e^2$ full width). The peak field amplitude on foil-2 is $9\times10^7$ V/m (corresponding to $1\times10^9$ W/cm$^2$) which induces an energy modulation with an amplitude of 1.8 eV, while the modulation amplitude on foil-1 is adjustable. An energy spectrum is obtained with a 2-sec exposure. We scan the delay time $\Delta t$ over ~7.5 fs with a 0.23-fs step. The data shown in this paper are the averages of 5-10 scans.

To attain attosecond resolution, the spread of the electron-light delay must be kept below a femtosecond. This is, however, not trivial for a beam with a large angular spread. On the $x$-$z$ plane (Fig. 1b), the travel distance between the foils changes with the angle $\alpha_{xz}$. For example, an electron with a trajectory toward the lower right ($\alpha_{xz} > 0$) travels a shorter distance than the one at $\alpha_{xz} = 0$ by $L\alpha_{xz}\tan\theta_{xz}$, where $\theta_{xz}$ is the angle of the foil (see Fig. 1b). In addition, due to the electron-light angle (90 degrees), the electron-light delay time also varies with the $x$-location on a foil where an electron passes through. Therefore, the vertical (i.e., along $x$) shift of the trajectory between the foils, approximately expressed as $L\alpha_{xz}$, alters the delay. An electron with $\alpha_{xz} > 0$ can interact with the light at an earlier time on foil-2 owing to the shift downwards. Accordingly, the choice of a proper foil angle $\theta_{xz}$ provides a uniform delay over the beam angles, as shown in Fig. 1d. The condition is expressed as $\tan\theta_{xz} = v_e/c$ when the laser beams are incident at right angles to the electron beam. For 37-keV electrons, this yields $\theta_{xz} = 20°$. The argument here holds regardless of the beam's focal position, but we spatially focus the beam on foil-2 such that there is no need to precisely align foil-2. On the *y*-*z* plane (Fig. 1c), the tilt angle of foil-1 ($\theta_{yz}$) is adjusted such that it is perpendicular to the beam, minimizing the spread of the flight times.

**Results and discussion**

Figure 2a shows an observed angle-resolved energy spectrum with the average delay time $\Delta t$. In addition to the zero-loss signals, signals attributed to plasmon loss appear at around a −20-eV shift. At the laser field strengths used here (1.9 eV amplitude for shaping and 1.8 eV for characterization), the zero-loss signals are well separated from those of the plasmon loss. Figure 2b shows the energy spectra at $\alpha_{yz} = -1.5\pm0.25$ mrad (upper panel) and $\alpha_{yz} = 0\pm0.25$ mrad (lower panel) as functions of the delay. Clear periodic modulations are observed in the spectra, indicating the successful production of electron pulses with sub-optical-cycle durations. The patterns are similar to those observed with accelerators [20,21]. When the two spectrograms are compared, a clear offset in the time axis is found. The offset becomes clearer when the center of mass (CoM) of the energy distribution is computed, as shown in Fig. 2d. They oscillate sinusoidally with time, and the signals at $\alpha_{yz} = -1.5$ mrad (blue squares) are delayed by ~0.2 fs with respect to the curve at $\alpha_{yz} = 0$ mrad (black circles). By repeating the same analysis for different $\alpha_{yz}$, we obtain the temporal offsets at different beam angles, as shown in Fig. 2c. The offset is nearly zero at small $|\alpha_{yz}|$ but becomes significant at $|\alpha_{yz}| > 1$ mrad. The difference in flight times shown in Fig. 1c alone (~0.05 fs at 2 mrad) cannot explain this result; we speculate that the relatively large offset originates from the curvature of the foil(s), which might be enhanced by the laser heating. The ability to observe the sub-femtosecond offsets demonstrates the high temporal resolution of our measurement.

The clear sinusoidal oscillations of the CoM reveal net acceleration at $\Delta t = T/4$, $5T/4$, and deceleration at $\Delta t = 3T/4$, $7T/4$, where $T = 3.4$ is the cycle time of the field. Moreover, the spectral intensities near zero shift become strongest (red color in Fig. 2b) at $\Delta t = T/2$ and $3T/2$. We note that the choice of the origin of $\Delta t$ is arbitrary; we define $\Delta t = 0$ as the delay time when the CoM of the energy distribution is zero and increasing. The enhancement in the spectral intensity suggests monochromatization of the beam. On the other hand, the weak intensities at $\Delta t = 0$, $T$ indicate broadening. The mechanism of broadening, acceleration, monochromatization, and deceleration controlled on sub-femtosecond timescales can be explained by considering the phase-space distribution of the optically modulated beam. As shown in the top part of Fig. 2g, the phase-space density right after foil-1 (blue curve) shows a sinusoidal oscillation. During propagation in vacuum from foil-1 to foil-2, the high-energy components catch up to the low-energy ones ahead of them, forming short bunches (blue vertical stripes in the lower parts). When the peak of the energy shift induced at foil-2 (red curve) coincides with the arrival of the bunches ($\Delta t = T/4$), net acceleration occurs (lowest part of Fig. 2g). Conversely, at a delay half a cycle later ($\Delta t = 3T/4$), the bunched electrons are decelerated. At the zero delay $\Delta t = 0$, the bunches experience the zero-crossing of the energy shift and undergo minimal change in their energy. Instead, the components outside the bunches whose energy increases with $z$ (upper right diagonal) are modulated. The sign of the energy shift (red curve) is the same as their relative energy, and accordingly the spectrum is broadened. On the contrary, at $\Delta t = T/2$, the opposite happens and monochromatization occurs. These characteristic sub-cycle features are well reproduced by a classical point-particle simulation (see Methods for details) shown in Fig. 2e.

The modulation field strength on foil-1 controls the temporal structure of the shaped electrons and the visibility of the attosecond streaking spectrogram accordingly. At about half the modulation strength ($5\times10^7$ V/m, 1.0-eV modulation), a similar pattern with a slightly weaker oscillation amplitude is observed (lower panel of Fig. 2h). A clear but very weak attosecond modulation is still observed at a strength as weak as $2\times10^7$ V/m (0.3-eV modulation) shown in the upper panel. To quantitatively evaluate the differences in the spectrograms, we evaluate the amplitude of the CoM oscillations. The black circles in Fig. 2i show the results; the CoM amplitude increases with stronger modulation. The net energy shift is mainly attributed to the acceleration or deceleration of the short bunches (Fig. 2g), thus the CoM amplitude can be used as a measure of the bunched pulse. When compared to the simulation (black curve in Fig. 2i), the experimental results (black circles) show the smaller CoM amplitudes. The potential origins are the aberration and astigmatism of the beam (i.e., more complex trajectories than shown in Figs. 1b-c.), vibration and thermal expansion of the foils, and the drift and jitter of the beam energy and the spectrometer during the scan. We incorporate these effects phenomenologically into our simulation as a fluctuation in the arrival time of the electrons at foil-2 ($\sigma_t$ in RMS), which effectively broadens the bunch length. At $\sigma_t = 0.4$ fs, the simulated CoM amplitudes (dotted curve) reproduce the experimental results well. At $\sigma_t = 0.6$ fs, the simulation results (dash-dotted line) completely underestimate the experimental ones. Figure. 2f shows the temporal profiles of the shaped electrons given by the simulations. At $\sigma_t = 0$ fs, the bunch duration is 0.5 fs in FWHM on a constant background, and 1.3 fs in FWHM with $\sigma_t = 0.4$ fs.

The bunch durations are evaluated in detail. Since the resolution of our spectrometer is insufficient to resolve photon-order sidebands, the direct reconstruction of temporal structures [18] is challenging. Therefore, we estimate the pulse durations based on models. First, using the point-particle model, we find $\sigma_t$ that yields good agreement with each experimental CoM amplitude. The black circles in Fig. 2j represent the FWHM pulse durations extracted from this analysis. Due to the relatively large $\sigma_t$ (~0.5 fs maximum) required for a good fit, the FWHM duration is nearly constant over the modulation strengths. Thus, the increase in the CoM amplitude in Fig. 2i corresponds to the enhancement of the pulse contrast compared to the time-constant background. Second, we conduct an analysis based on a simpler model. We assume a Gaussian pulse shape on a constant background and find the pulse width for the best match with the experimental result (see Methods for details). The results in black squares show a decrease in the bunch duration with modulation strength. We note that this latter model tends to overestimate the pulse width as it approaches a cycle period (3.4 fs), because the width is defined without accounting for the overlap between adjacent pulses (i.e., pulses before and after a cycle). The results of both analyses are consistent at the modulation amplitudes larger than 1.5 eV and give the shortest duration of 1.3±0.2 fs in FWHM and 0.5±0.1 fs in RMS at the modulation amplitudes of 1.8-1.9 eV.

Finally, we report on the attosecond modulation of electron beams whose energy spreads are significantly broadened due to the space charge effect. Temporal compression may not occur effectively in beams with broad energy distributions due to the inherent velocity spread. The two spectrograms shown in Fig. 3a are observed at almost the same laser field strengths as above (i.e., 2.0 and 1.8 eV amplitudes at foil-1 and foil-2, respectively). The upper panel is the spectrogram observed at a 15-eV (FWHM) spread with ~16 electrons/pulse (without slit) while that in the lower panel corresponds to the beam of a 9-eV spread with ~10 electrons/pulse. Although the modulations in these images are subtle because the energy bandwidth is much larger than the light-induced energy shifts, clear CoM oscillations still appear, as shown in Fig. 3b, suggesting the production of pulses of sub-cycle durations. The CoM oscillation amplitudes are 0.09 (black circles) and 0.27 eV (blue squares), respectively. For comparison, we simulate the temporal profiles and CoM amplitudes using the point-particle model with $\sigma_t$ = 0.4 fs. We assume no energy-time correlation (e.g., chirp). Figure 3c shows the simulated temporal structures of the shaped beams. As the energy spread increases, the peak profiles gradually fade away and become almost invisible at a spread of 10 eV (FWHM). In Fig. 3d, both the experimental (circles) and simulated (curve) CoM amplitudes decay with the increase of energy spread. However, the simulated curve underestimates the experimental results. The amplitudes of 0.09 and 0.27 eV are given by simulations with smaller energy spreads of 8 and 5 eV, respectively, whose temporal profiles in Fig. 3c still hold sub-cycle peak structures. We speculate that this difference is attributed to the chirp of the energy-broadened beam. Although the entire beam has a wide energy spread, the instantaneous spread on the time scale of an optical cycle period is much narrower, and accordingly the optical shaping works.

**Summary and conclusion**

In conclusion, we developed the electron beamline at the energy of 37 keV and demonstrated attosecond beam shaping with light at a high beam current. With a pulse containing 2-electrons (without the slit) in the 290-fs envelope, we produced and measured a train of pulses with durations of 1.3±0.2 fs (FWHM) and 0.5±0.1 fs (RMS) at the modulation amplitudes of 1.8-1.9 eV. In addition, we observed clear sub-cycle signatures in the spectrograms at higher currents up to 16 electrons/pulse with a modulation amplitude of 2 eV. The slit was used only for energy analysis, and the flux reaching a sample can be enhanced by placing it behind the foils. With stronger modulation with a longer cycle period (e.g. mid-infrared or far-infrared light) [16], generation of attosecond pulses with more than one electron/micro-pulse is within reach. The use of a pre-buncher or a monochromator using THz [45–47] would also be helpful in achieving this. The sub-femtosecond electron pulses with high average currents produced in this study could facilitate attosecond diffractive imaging and controlled photon generation.

## Methods

### Point-particle simulation details

The simulations are conducted in one dimension. A group of electrons with random kinetic energies is prepared to reproduce the bandwidth. For electron beams with negligibly small space charge effects, a 1.2-eV bandwidth (FWHM) is assumed. These electrons are then subjected to time-periodic acceleration and deceleration, determined by their initial positions. The electrons propagate over a distance of 2.6 mm before arriving at foil-2. To phenomenologically incorporate the effects of jitter as well as the energy-time uncertainly, the arrival times are randomly shifted according to a normal distribution with an RMS width given by the sum of $\sigma_t$ and $2\hbar/\Delta E$. Here $\sigma_t$ is the only free parameter in this model and $\Delta E$ denotes the RMS width of the energy spectrum after modulation. The temporal structure of the electrons is obtained from the arrival times. A streaking spectrogram is calculated by considering the additional time-periodic energy shifts determined by the arrival times.

### Simple model for pulse width estimation

Because all half-cycles of the modulation field do not contribute to the temporal pulse compression, we assume that half of the total electrons form pulses with a Gaussian shape, while the other half constitutes the time-independent background. By assuming that all electrons have the same kinetic energy before the energy modulation at foil-2, the center of mass of the energy distribution is uniquely determined by the pulse duration. The energy modulation at foil-1 and the propagation effect are not considered in this model. Because the amplitude of the CoM oscillation is given by the proportion of electrons within a finite time interval during which acceleration or deceleration occurs, this model can provide a good estimate. When the pulse width is long, the increase in the constant background due to the overlap with neighboring peaks tends to make the actual time width shorter than the model's value.

## Acknowledgement

This research was supported by MEXT/JSPS KAKENHI JP21K21344, JP25K22230, JP 25K01734, JST FOREST JPMJFR2228 and ACT-X JPMJAX21AO, Gordon and Betty Moore Foundation, Kazato Research Foundation, Research Foundation for Opto-Science and Technology, Yamada Science Foundation. We would like to thank Advanced Manufacturing Support Team at RIKEN for their support in design and manufacturing of components. Y.M. acknowledges Tomoya Okino and Yuichiro Kato for their assistance in setting up the laboratory.

**Author contributions**

Y.M. conceived the study. T.H., Yo.T., Y.Y. and Y.M. designed and constructed the electron gun. Yu.T. and Y.M. designed and developed the electron energy spectrometer. M.O. and Y.M. set up the optical system. Yu.T. and T.I. developed programs for automatic data collection. Yu.T. and Y.M. performed the experiments and analyzed the data. Y.M. wrote the manuscript with input from all authors.

**Data availability statement**

The datasets analyzed during the current study are available from the corresponding authors on reasonable request.

**Ethics declarations**

The authors declare no conflicts of interest associated with this manuscript.

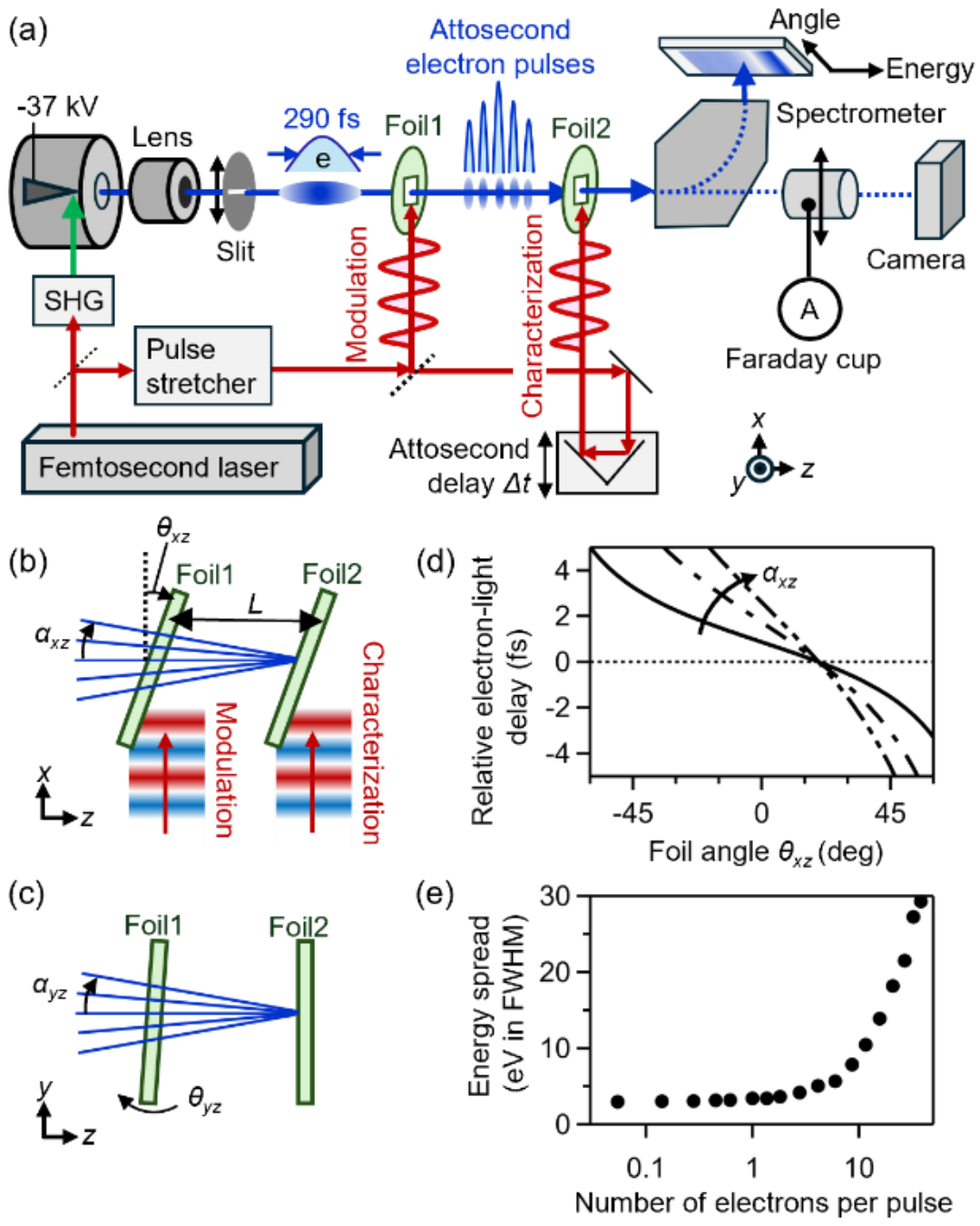

**Fig.1. Attosecond shaping experiment in a home-built electron beamline.** (a) Schematic of the experimental setup. See the text for details. (b) Arrangement of foils on the *x*-*z* plane (green) to minimize the spread in light-electron delays. An appropriate tilt angle (20° for 37-keV electrons) yields a light-electron delay time distribution narrower than a femtosecond over a wide range of incident angles. (c) Arrangement on the *y*-*z* plane. (d) Delay times between electrons and the laser compared to $\alpha_{xz} = 0$ mrad. Solid, dash-dotted, and dash-dot-dotted lines corresponds to $\alpha_{xz} = 0.1$, 0.2 and 0.3 mrad, respectively. (e) Energy spread due to the space charge effect. The resolution of the spectrometer is 3 eV (FWHM).

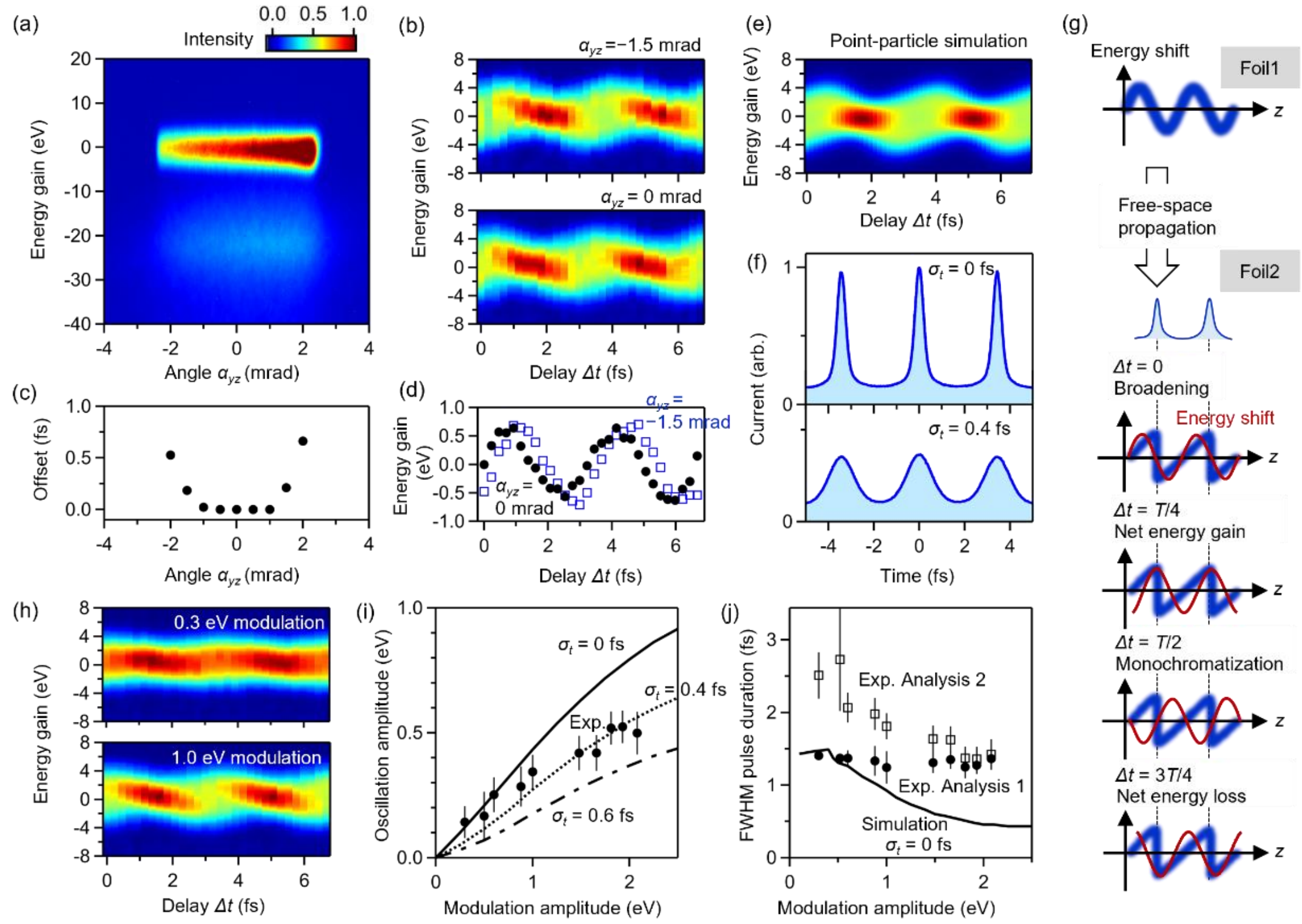

**Fig. 2. Attosecond modulation and characterization results.** (a) Raw image of the signals. (b) Attosecond spectrograms obtained at the beam angles of $\alpha_{yz} = -1.5$ mrad (upper panel) and $\alpha_{yz} = 0$ (lower panel). (c) Temporal offsets at different beam angles. (d) Center-of-mass (CoM) analysis of the delay-dependent energy shift. (e) Attosecond streaking spectrogram simulated with the point-particle model, see Methods. (f) Simulated temporal densities with jitters of $\sigma_t = 0$ fs and $\sigma_t = 0.4$ fs. (g) Schematic diagrams illustrating the origins of the spectral changes caused by sub-femtosecond delays. (h) Spectrograms observed with weaker modulation amplitudes. Signals are integrated at the angles from $\alpha_{yz} = -0.5$ to $\alpha_{yz} = +0.5$ mrad. (i) Comparison of the observed CoM amplitudes (black circles) with simulation results (curves). (j) Extracted FWHM pulse durations. See the text for details.

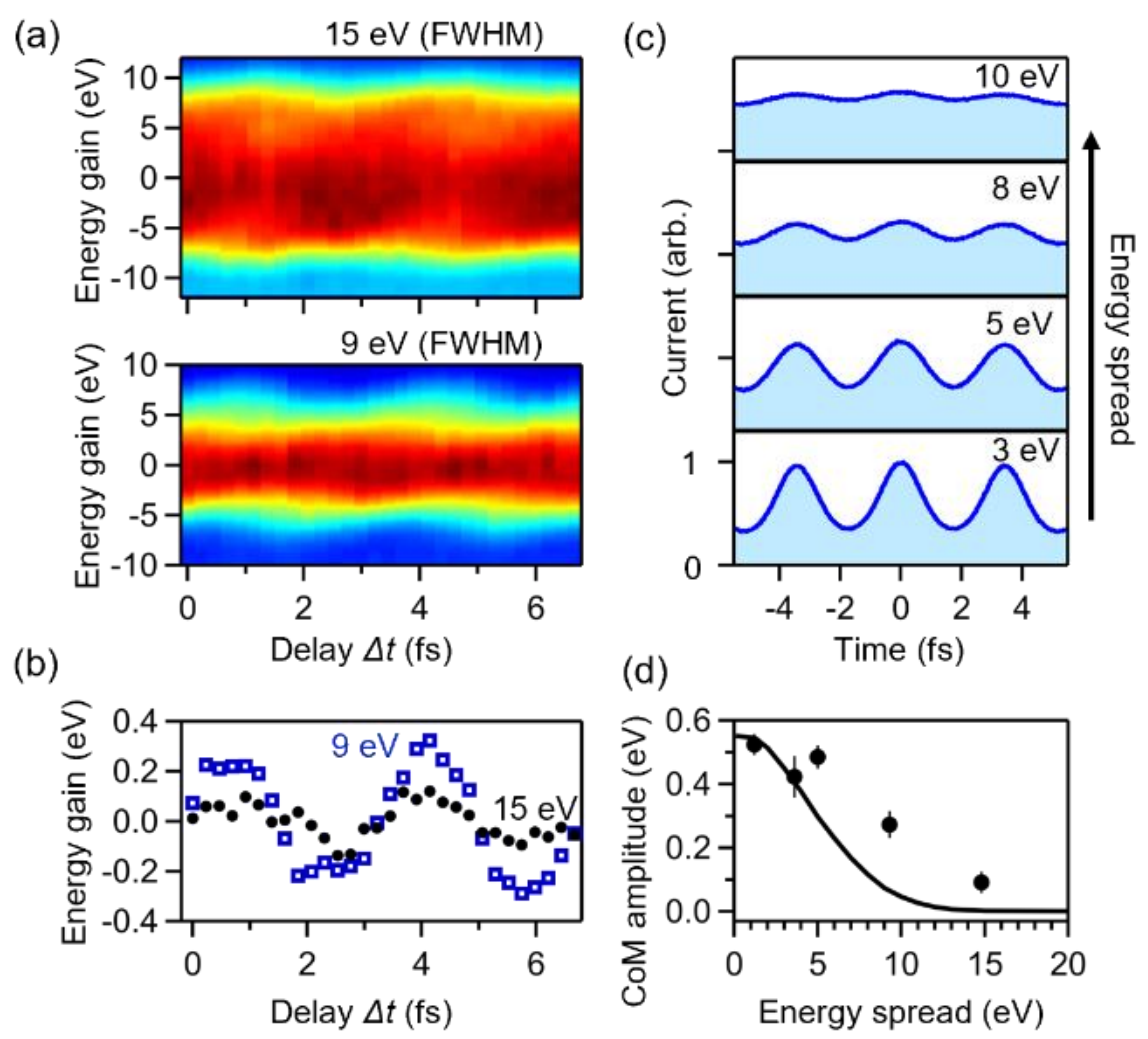

**Fig. 3. Sub-optical-cycle modulation of electron beams with strong space-charge effects.** (a) Observed spectrograms with the electron beams having the energy spreads of 15 eV (upper panel) and 9 eV (lower panel). The probe currents without the slit are 16 and 10 electrons/pulse, respectively, at the sample location. Signals are integrated at angles from $\alpha_{yz} = -0.5$ to +0.5 mrad. (b) Observed CoM oscillations as a function of attosecond delays. Squares and circles represent the results for the energy spreads of 9 and 15 eV. (c) Simulated temporal densities of the shaped electrons with different levels of energy spread. The modulation amplitude is 2 eV. (d) Comparison of the experimentally obtained CoM oscillation amplitudes (circles) with the simulation results (curve). The amplitudes larger than the simulations indicate the beam chirp and shorter bunch lengths.